\documentclass[final,5p,times]{elsarticle}
\usepackage{lineno}
\usepackage{url}
\usepackage{graphicx}
\usepackage{amstext,amsmath,amssymb,amsthm}
\usepackage{upgreek}
\usepackage{enumerate}
\usepackage{txfonts}
\biboptions{sort&compress}
\journal{Nuclear Instruments and Methods A}

\begin{document}

\makeatletter
\def\@cite#1{[{#1}]}
\makeatother

\begin{frontmatter}

\title{A novel analog power supply for gain control of the Multi-Pixel Photon Counter (MPPC)}

\author[IHEP]{Zhengwei Li\corref{cor1}}\ead{lizw@ihep.ac.cn}
\author[IHEP]{Congzhan Liu}
\author[IHEP]{Yupeng Xu}
\author[IHEP,UCAS]{Bo Yan}
\author[IHEP]{Yanguo Li}
\author[IHEP]{Xuefeng Lu}
\author[IHEP]{Xufang Li}
\author[IHEP,UCAS]{Shuo Zhang}
\author[IHEP,UCAS]{Zhi Chang}
\author[IHEP,YNU]{Jicheng Li}
\author[IHEP,UCAS]{He Gao}
\author[IHEP]{Yifei Zhang}
\author[IHEP]{Jianling Zhao}
\cortext[cor1]{Corresponding author. }

\address[IHEP]{Key Laboratory of Particle Astrophysics, Institute of High Energy Physics, Chinese Academy of Sciences, Beijing, China}
\address[UCAS]{University of Chinese Academy of Sciences, Beijing, China}
\address[YNU]{Yunnan University, Kunming, China}

\begin{abstract}

Silicon Photo-Multipliers (SiPM) are regarded as novel photo-detectors to replace conventional Photo-Multiplier Tubes (PMTs). However, the breakdown voltage dependence on the ambient temperature results in a gain variation of $\sim$3$\% /^{\circ} \mathrm C$. This severely limits the application of this device in experiments with wide range of operating temperature, especially in space missions. An experimental setup was established to investigate the temperature and bias voltage dependence of gain for the Multi-Pixel Photon Counter (MPPC). The gain and breakdown voltage dependence on operating temperature of an MPPC can be approximated by a linear function, which is similar to the behavior of a zener diode. The measured temperature coefficient of the breakdown voltage is $(59.4 \pm 0.4$ mV)$/^{\circ} \mathrm C$. According to this fact, an analog power supply based on two zener diodes and an operational amplifier was designed with a positive temperature coefficient. The measured temperature dependence for the designed power supply is between 63.65 to 64.61~mV/$^{\circ} \mathrm C$ at different output voltages. The designed power supply can bias the MPPC at an over-voltage with a temperature variation of $\sim$ 5~mV$/^{\circ} \mathrm C$. The gain variation of the MPPC biased at over-voltage of 2~V was reduced from 2.8$\% /^{\circ} \mathrm C$ to 0.3$\% /^{\circ} \mathrm C$ when biased the MPPC with the designed power supply for gain control. Detailed design and performance of the analog power supply in the temperature range from -42.7$^{\circ}\mathrm{C}$  to 20.9$^{\circ}\mathrm{C}$  will be discussed in this paper.

\end{abstract}


\end{frontmatter}


\section{Introduction}
The Silicon Photomultiplier~(SiPM) has become a promising device for applications in medical imaging, high energy particle physics and particle astrophysics with high Photon Detection Efficiency~(PDE), high gain (up to $10^{6}$), low cost, low operating voltage~($<100\,\mathrm{V}$), excellent timing resolution~($\sim 120\,\mathrm{ps}$) and insensitivity to magnetic field \cite{Lu201430,Buzhan2003,Kovaltchouk2005,DelGuerra2011}. The SiPM also referred to as SSPM, AMPD, MPPC, MRSAPD. It consists of an array of Avalanche Photodiodes~(APDs) working in the Geiger mode. The APDs are biased above the breakdown voltage $(V_ \mathrm{BD})$ with quenching resistor in serial and connected in parallel. The difference between the bias voltage $(V_\mathrm{bias})$ and breakdown voltage is called over-voltage $(\Delta V=V_\mathrm{bias}-V_\mathrm{BD})$.

Their insensitivity to magnetic field, low operating volatge ($<100$~V), compact size and sensitivity to a small number of photons make them novel light detectors for the scintillator detectors on board the Hard X-ray Modulation telescope (HXMT)~\cite{Li201663}. One of the most important problems for the application of SiPM onboard the HXMT is the gain dependence on the operating conditions (bias voltage, operating temperature). The gain of SiPM is a linear function of the over-voltage. The breakdown voltage has a positive temperature coefficient about $50\mathrm{mV}/^{\circ}\mathrm C$ which results in a typical gain variation of $\sim$3$\% /^{\circ} \mathrm C$. The scintillator detectors on board the HXMT will operate in the temperature range from -40$^{\circ}\mathrm{C}$  to 15$^{\circ}\mathrm{C}$. Therefore, it is essential to design a gain control and stabilization system for the SiPM to operate in a wide range of temperatures.

One way to keep the gain of a SiPM stable is to maintain the operating temperature at a constant value by using external temperature control system. This method is mainly used in ground experiments, such as the outer hadron calorimeter of the CMS detector~\cite{Freeman2010393}. The temperature control system need a cooling and heating system to keep the temperature stable, which is not suitable for the scintillator detectors onboard HXMT. As the extra power budget for the cooling and heating system is unacceptable for the HXMT. An alternative approach to maintain a stable gain is to bias the SiPM at a constant over-voltage. The bias voltage applied to the SiPM should be adjusted according the environment temperature. This can be implemented in different ways such as closed loop feedback of the dark current ~\cite{2012lee}, use of thermistors to compensate the drift of the over voltage~\cite{Miyamoto2009}, a linear feedback on operating bias voltage~\cite{Marrocchesi2009,Dorosz2013202,Baszczyk201685,Shukla} or variable gain amplifiers to compensate the variation of the SiPM gain~\cite{Yamamoto2011}. Even the SiPM can be used as a temperature sensor according to the amplitude of dark noise pulses to control the gain. The gain of SiPM can be measured as a function of temperature according to the average amplitude of the dark noise pulses. A negative feedback loop is established to adjust the bias voltage of the SiPM according to the variation of the gain to keep the gain stable~\cite{Licciulli2013}. However, the method that utilizing the dark current or thermistors would induce a large gain deviation from the constant value when operating in a large temperature range. This limits its application in the case of HXMT. Secondly, linear feedback or negative feedback on operating voltage need the design of a complex feedback electronic system based on digital processing unit, which limits its use in the HXMT. The method that adjust the SiPM bias return line according the temperature-to-voltage converter~\cite{Bencardino2009} is simple enough to be used in the HXMT. This method needs a precise power source to bias the SiPM at a fixed voltage. In the case of HXMT, a space qualified precise power source is needed. This limits the use of such method onboard the HXMT.

In this paper, a simple analog power supply system for gain control and stabilization of MPPC on board HXMT was developed by using zener diode as temperature sensor. The MPPC is one type of UV sensitive SiPM with a p-on-n structure developed by Hamamatsu with relative low dark count rate~\cite{Musienko2009a,Danilov2009}. The type of MPPC used on board HXMT is S10362-33-050C, consisting of 3600 pixels, covering geometrically $61.5\%$ of total area of $3\mathrm{mm}\times 3\mathrm{mm}$. The breakdown voltage of zener diode has a positive/negative linear temperature coefficient according to its breakdown voltage. An analog power supply system with a linear positive temperature coefficient can be designed based on zener diodes. Different values of temperature coefficient can be obtained by using different zener diodes with different breakdown voltage. The designed analog power supply system is a closed loop feedback system to bias the SiPM at a fixed over-voltage at different environment temperature by adjusting the bias voltage automatically.

The working principle of the analog power supply system was presented in Section~\ref{sect:principle}. In Section~\ref{sect:temperature}, the temperature dependence of the MPPC, zener diodes and designed analog power supply system were investigated.
In Section~\ref{sect:performance}, the performance of the MPPC biased by the designed analog power supply, has been investigated for demonstration of gain stability in the temperature range from -42.7$^{\circ}\mathrm{C}$  to 20.9$^{\circ}\mathrm{C}$. The measured results show that the designed power supply system for gain control of MPPC meets the operating temperature requirement on board HXMT. A further optimized simple analog circuit to maintain the gain of a large array of MPPC was proposed in Section~\ref{sect:discussion}.

As the designed analog power supply system will be used in outer space, the zener diodes should be space-qualified product. However, there are no significant difference between the measured temperature coefficients of the zener voltage for the space-qualified products and industrial products. So the results of the space-qualified product is not shown in this paper.

\section{Working principle of the analog power supply}\label{sect:principle}
 The MPPC consists of an array of avalanche photo-diodes operating in Geiger mode with the diodes reversely biased above the electrical breakdown voltage ($V_\mathrm{BD}$). At this bias voltage, the electric field in the diode depletion region is sufficiently high for free carriers in the depletion region to produce additional carriers by impact ionization, resulting in a self-sustaining avalanche. The free carriers can be generated by incident photons or thermal emission.
 The total charge in the avalanche can be evaluated as $Q=C_\mathrm{pixel}\times \Delta V$, where $C_\mathrm{pixel}$ is the effective pixel capacitance. Neglecting the capacitance difference among the different pixel cells, the integrated charge is identical for different cells. The MPPC gain is defined as the charge produced in a single pixel avalanche, expressed in elementary charge unit
\begin{equation}
\label{eq:gaindeltav}
G(\Delta V)=\frac{Q}{e}=\frac{C_\mathrm{pixel}}{e}(V_{\mathrm{bias}}-V_{\rm{BD}})=\frac{C_\mathrm{pixel}}{e}(V_{\mathrm{bias}}-V^{0}_{\rm{BD}}-\alpha T).
\end{equation}
\begin{figure}[!htp]
\begin{center}
\includegraphics[width=0.9\linewidth,clip]{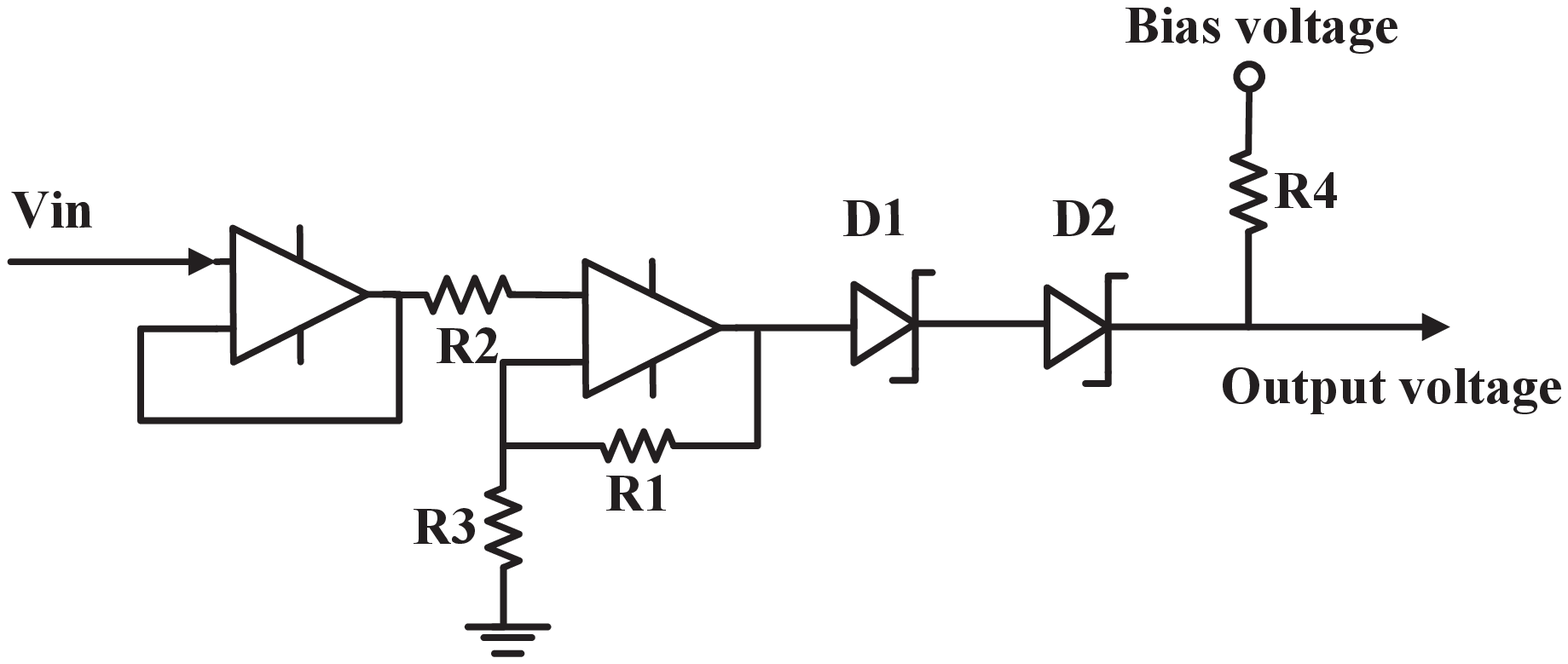}
\caption{Block diagram of the analog power supply. The bias voltage applied to the resistor R4 is used to keep the zener diodes working in avalanche mode. The output voltage of the power supply can be controlled by the input voltage $V_{\mathrm in}$.}
\label{fig.voltagesource}
\end{center}
\end{figure}
The parameter $\alpha$ in Equation~(\ref{eq:gaindeltav}) represents the temperature coefficient of the breakdown voltage of the MPPC, and the parameter $V^{0}_{\rm{BD}}$ represents breakdown voltage of MPPC at the temperature of 0$^{\circ} \mathrm C$. Therefore, the gain of MPPC could be maintained at a constant value if the MPPC is biased by a voltage source with a positive temperature coefficient of $\alpha$.

A simple analog power supply system based on two zener diodes and a dual operational amplifier is designed, as shown in Figure.~\ref{fig.voltagesource}. The voltage between the resistor R4 and zener diode D2 is referred as the output voltage ($V_{\mathrm out}$). The output voltage is connected to the MPPC to bias the MPPC at voltage of $V_{\mathrm out}$. A programmable voltage $V_{\mathrm in}$ was used as the input voltage to control the output voltage $V_{\mathrm out}$, which can be expressed as flowing
\begin{equation}
  V_{\mathrm out}(V_{\mathrm in},T) = A \cdot V_{\mathrm in}+V_{\mathrm zener}(\mathrm D1)+V_{\mathrm zener}(\mathrm D2).
\label{eq:outputvoltage}
\end{equation}

The parameter A in Equation~\ref{eq:outputvoltage} can be expressed as $A= \frac{\mathrm R3+\mathrm R1}{\mathrm R3}$.
When the zener diode is working in the avalanche mode, the zener voltage has a positive linear temperature coefficient. From the equation~(\ref{eq:outputvoltage}), it can be found that the output voltage $V_{\mathrm out}$ have a temperature coefficient which equals to the sum of that of the two zener diodes. This indicates that an analog power supply  with different temperature coefficient can be obtained by using zener diodes with different temperature coefficient. To set the two zener diodes working in avalanche mode, a bias voltage should be applied to the resistor R4. The applied bias voltage should be larger than the output voltage $V_{\mathrm out}$. In this paper, the bias voltage is about 80V and the designed value of R4 is about 50~$\mathrm{k}\Omega$. The zener voltage of the D1 and D2 is approximately 33~V, so the current that flowing through D2 and D1 is larger than 200~$\mu$A. If the bias voltage changes about 2V, the current that flowing through the D2 and D1 will change less than 40uA. This change is small compared to the 200uA current. And the zener diodes is working in the avalanching mode, the change of the zener voltage that across the Zener diodes induced by the variance of flowing current could be neglected. This indicates that the $V_{\mathrm out}$ is independent on the bias voltage which is applied to the resistor R4, just as shown in equation~(\ref{eq:outputvoltage}).

The independence on the bias voltage reduces the requirement for a specific precise power supply system. A simple charge-pump circuit that regulate the 12~V to 80~V can meet the requirement of the power supply to bias the resistor R4 at approximately 80~V. As the $V_{\mathrm out}$ is independent on the bias voltage, the temperature dependence of the charge-pump circuit will not effect the performance of output voltage $V_{\mathrm out}$. This simplifies the design of the MPPC readout electronics.

The over-voltage of MPPC biased by the analog power supply can be expressed as flowing
\begin{equation}
\begin{split}
& \Delta V(V_{\mathrm in},T) = A \cdot V_{\mathrm in}+V_{\mathrm zener}(\mathrm D1)+V_{\mathrm zener}(\mathrm D2)-V_{\mathrm BD} \\
& =A \cdot V_{\mathrm in}+(V^{0}_{\mathrm zener}(\mathrm D1)+V^{0}_{\mathrm zener}(\mathrm D2)-V^{0}_{\mathrm BD})+(\alpha_{1}+\alpha_{2}-\alpha)\cdot \mathrm T \\
\end{split}
\label{eq:overvoltage}
\end{equation}
where $\alpha_{1}$ and $\alpha_{2}$ represents the zener voltage temperature coefficient of diode D1 and D2 respectively, $V^{0}_{\mathrm zener}$ represents the zener voltage of the diode at $0^{\circ} C$. If the sum of the zener voltage temperature coefficients of the two zener diodes equals that of the MPPC, the over-voltage of MPPC can be maintained at a constant value. This means that the gain of MPPC can be kept at a constant value at different environment temperatures.
\section{Temperature dependence}\label{sect:temperature}
\begin{figure}[htp]
\begin{center}
\includegraphics[width=1.0\linewidth,clip]{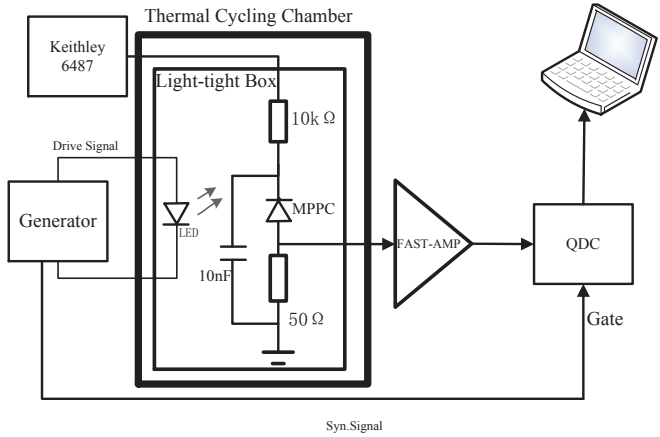}
\caption{Experimental setup.}
\label{fig.setup}
\end{center}
\end{figure}
An experimental setup in dark condition was established to investigate the temperature and bias voltage dependence of the gain for the MPPC. The temperature coefficient of the breakdown voltage of the MPPC could be derived from the gain dependence on operating temperature and bias voltage. The derived temperature coefficient is used as the baseline specification of the temperature coefficient for the design of the analog power supply system. According to the measured temperature coefficient of MPPC, the zener diode with a temperature coefficient of about 30~mV$/^{\circ} \mathrm C$ was chosen as the zener diode of the power supply system.

This method can be generalized to other MPPC and SiPM with different temperature coefficient. The temperature coefficient of zener diode working in different zener voltage is different, it increases with the zener voltage of the diode. The temperature coefficient of the BZV55 series zener voltage ranges from -3.5~mV$/^{\circ} \mathrm C$ to 88.6~mV$/^{\circ} \mathrm C$. So, if the temperature coefficient of the SiPM is different, different zener diodes with different coefficient could be used in series to bias the SiPM. The total temperature coefficient of the zener diodes in series is the sum of that all of the zener diodes. In this work the BZV55-C33 zener diode is chosen, with a typical temperature coefficient ranging from 23.3~mV$/^{\circ} \mathrm C$ to 33.4~mV$/^{\circ} \mathrm C$ according to the datasheet of zener diode. 14 samples of BZV55-C33 were measured in this paper. It is found that the ture temperature coefficients are almost the same for the 14 samples , ranging from 31.31~mV$/^{\circ} \mathrm C$ to 32.18~mV$/^{\circ} \mathrm C$. The detailed performance of the zener diode and designed analog power supply system were investigated.
\subsection{Experimental setup}\label{sect:setup}
\begin{figure}[htp]
\begin{center}
\includegraphics[width=0.8\linewidth,clip]{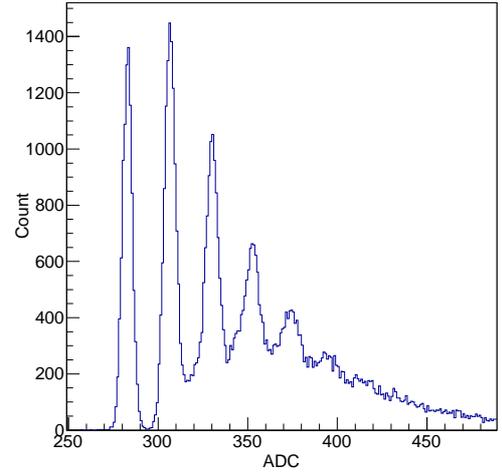}
\caption{Measured pulse-height distribution of the MPPC illuminated by the LED.}
\label{fig.pulseheight}
\end{center}
\end{figure}
The experimental setup shown in Fig.~\ref{fig.setup} has been developed to study the characterization of the MPPC. A fast LED driven by the generator was used as light source to illuminate the MPPC. The bias voltage of the MPPC is provided by a Keithley 6487 Picoammeter/Voltage Source. The signal from the MPPC is amplified by a fast amplifier (designed with the OPA 657, gain G=-20). And then the amplified signal is readout by a QDC (CAEN V965A) with 33ns gate width. The 33ns gate signal was the synchronizing signal of the LED drive signal which was provided by the generator. The biased circuit of the MPPC and the LED were placed in a light-tight box. The light-tight box was placed into a climatic chamber to study the performance of MPPC at different environment temperatures.
\begin{figure*}[htp]
\begin{center}
\includegraphics[width=0.35\linewidth,clip]{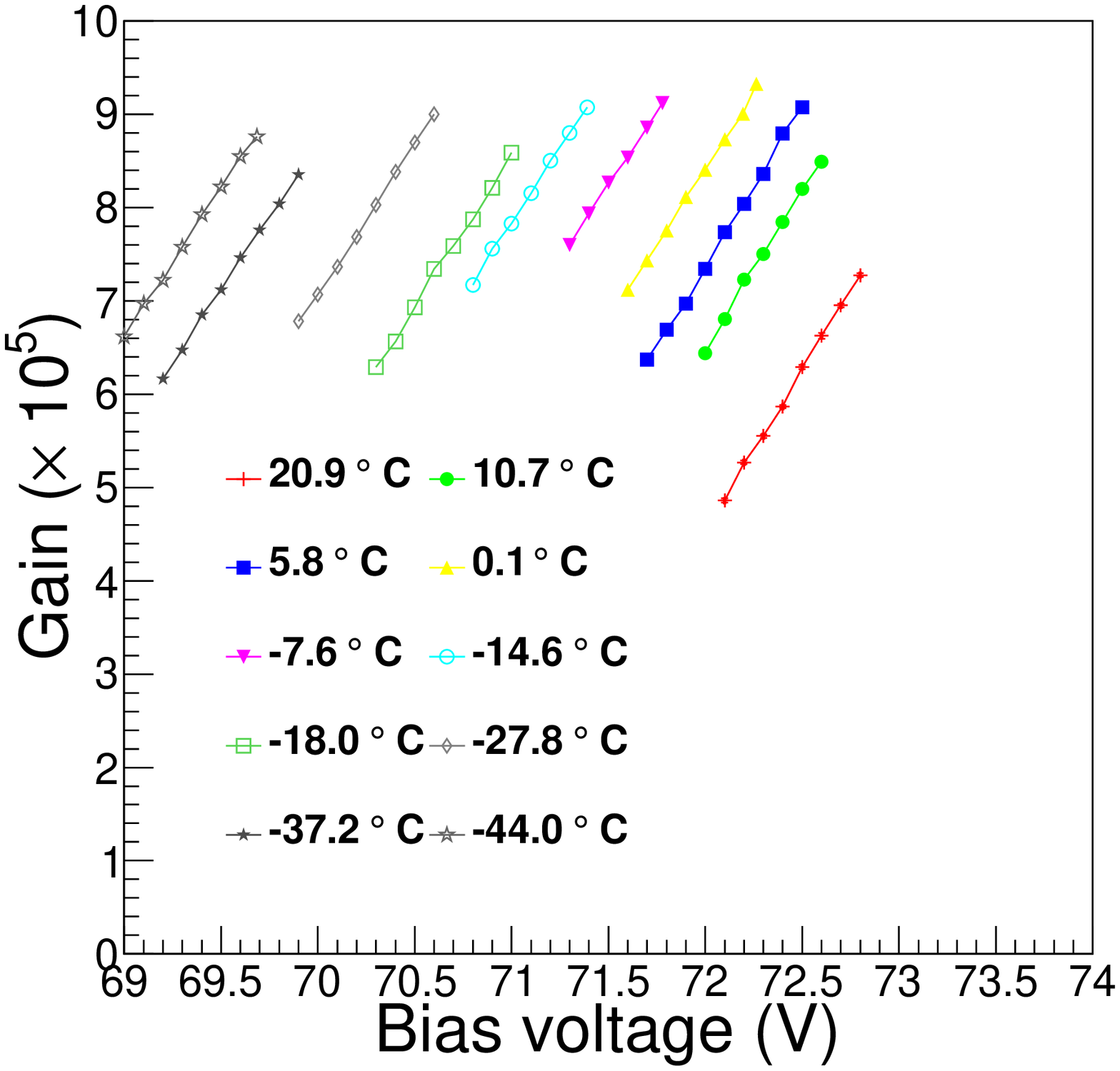}
\includegraphics[width=0.35\linewidth,clip]{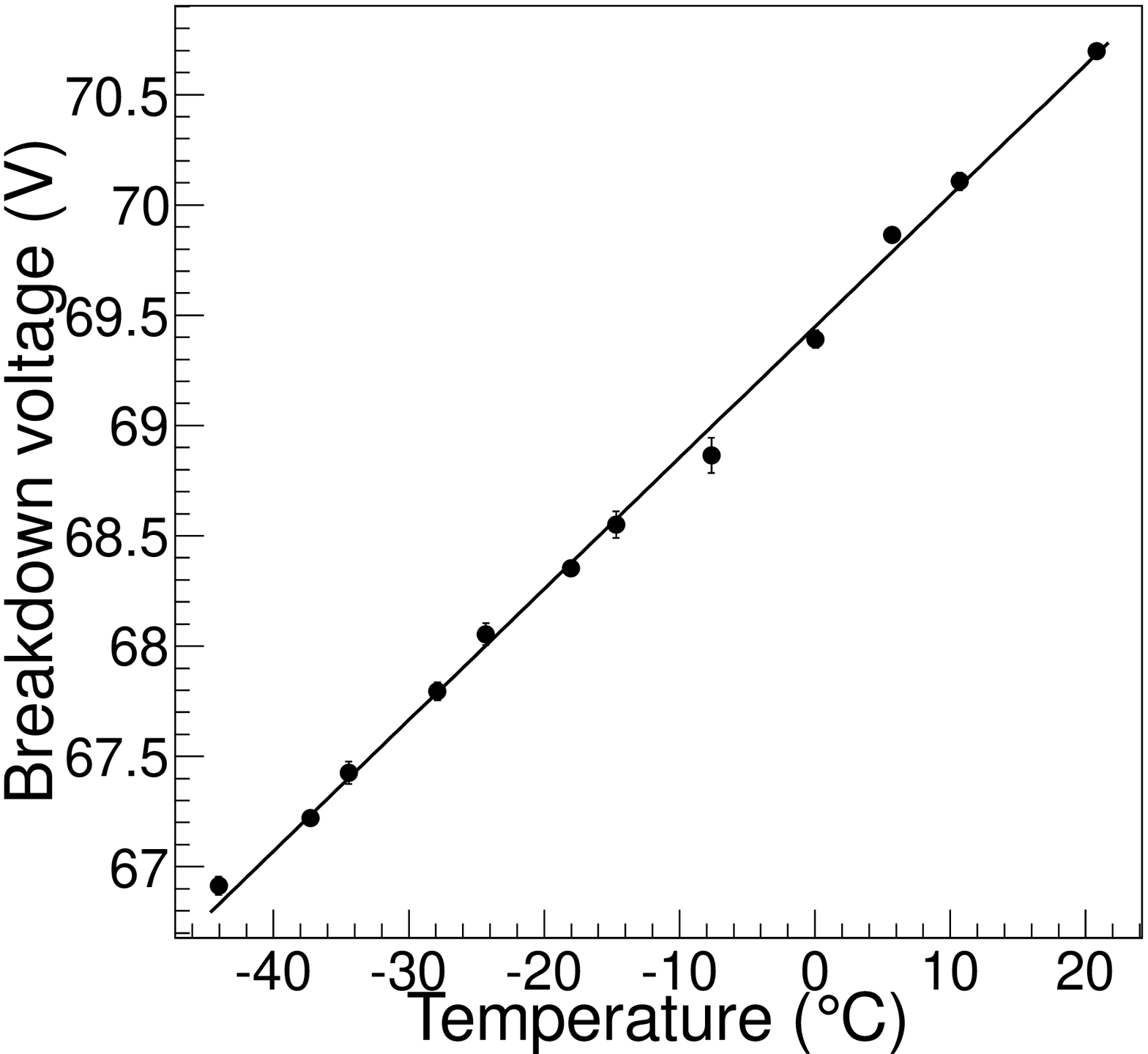}
\caption{Left panel: Measured gain of MPPC as function of bias voltage in the temperature range from $-44^{\circ}\mathrm{C}$ to $20.9^{\circ}\mathrm{C}$. Right panel: Measured breakdown voltage of MPPC as function of operating temperature. The linear line shown in the right panel is the linear fitted result.}
\label{fig.breakdownvoltage}
\end{center}
\end{figure*}
\subsection{Gain and breakdown voltage dependence of the MPPC}
\begin{figure}[htp]
\begin{center}
\includegraphics[width=0.8\linewidth,clip]{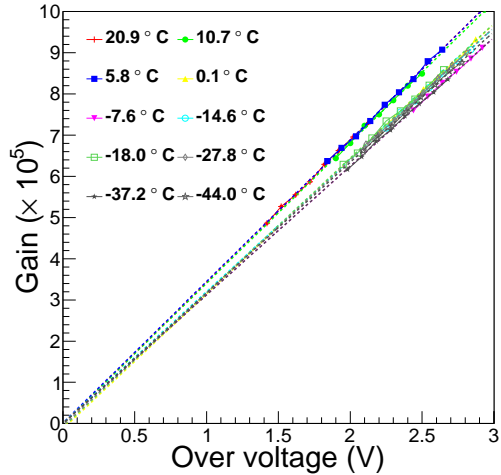}
\caption{Measured gain vs over-voltage for different operating temperatures. The lines shown in the figure are the linear fitting results.}
\label{fig.gainvsovervoltage}
\end{center}
\end{figure}
The gain and breakdown voltage dependence of the MPPC were investigated by using the experimental setup as shown in Fig.\ref{fig.setup}. The light intensity of the LED that illuminated the MPPC was tuned to make sure that the number of photons incident on to the sensitive area of the MPPC was below 10. In this way, the single equivalent photon pulse-height distribution can be measured with clearly discrete peaks, just as shown in Fig.~\ref{fig.pulseheight}. The distance between the different single equivalent photon peaks is linearly proportional to the gain of the MPPC. By dividing this distance with the conversion gain of the QDC and the electron charge, the gain of the MPPC can be measured.

The measured gain of the MPPC as a function of bias voltage is shown in Fig.~\ref{fig.breakdownvoltage}. At a fixed operating temperature, the gain of the MPPC is a linear function of bias voltage applied to the MPPC. By fitting the gain dependence on the bias voltage with linear equation~(\ref{eq:gaindeltav}), the breakdown voltage of the MPPC can be derived at a given operating temperature. The measured breakdown voltage of the MPPC in the temperature range from $-44^{\circ}\mathrm{C}$ to $20.9^{\circ}\mathrm{C}$ was shown in Fig.~\ref{fig.breakdownvoltage}. The breakdown voltage of the MPPC increases as a linear function of operating temperature. The fitted linear temperature coefficient of the breakdown voltage from Fig.~\ref{fig.breakdownvoltage} is $(59.4 \pm 0.4$ mV)$/^{\circ} \mathrm C$. This indicates that the analog power supply connected to the MPPC should have a positive temperature coefficient of $59.4 \mathrm mV/^{\circ} \mathrm C$ to keep the gain at a constant value at different operating temperatures.

The measured gain dependence on the over-voltage is shown in Figure.~\ref{fig.gainvsovervoltage}. It was derived from Figure.~\ref{fig.breakdownvoltage} by subtracting the breakdown voltage from the bias voltage. It is found that all of the linear curves of the gain dependence on the over-voltage passed through the zero origin point. The slope of the linear function becomes slightly smaller as temperature decreases. This will result in an deviation from constant value for the gain of MPPC even if the bias voltage of the MPPC is adjusted correctly according to the temperature coefficient of the MPPC.

\subsection{Breakdown voltage dependence of the zener diode}
\begin{figure*}[htp]
\begin{center}
\includegraphics[width=0.35\linewidth,clip]{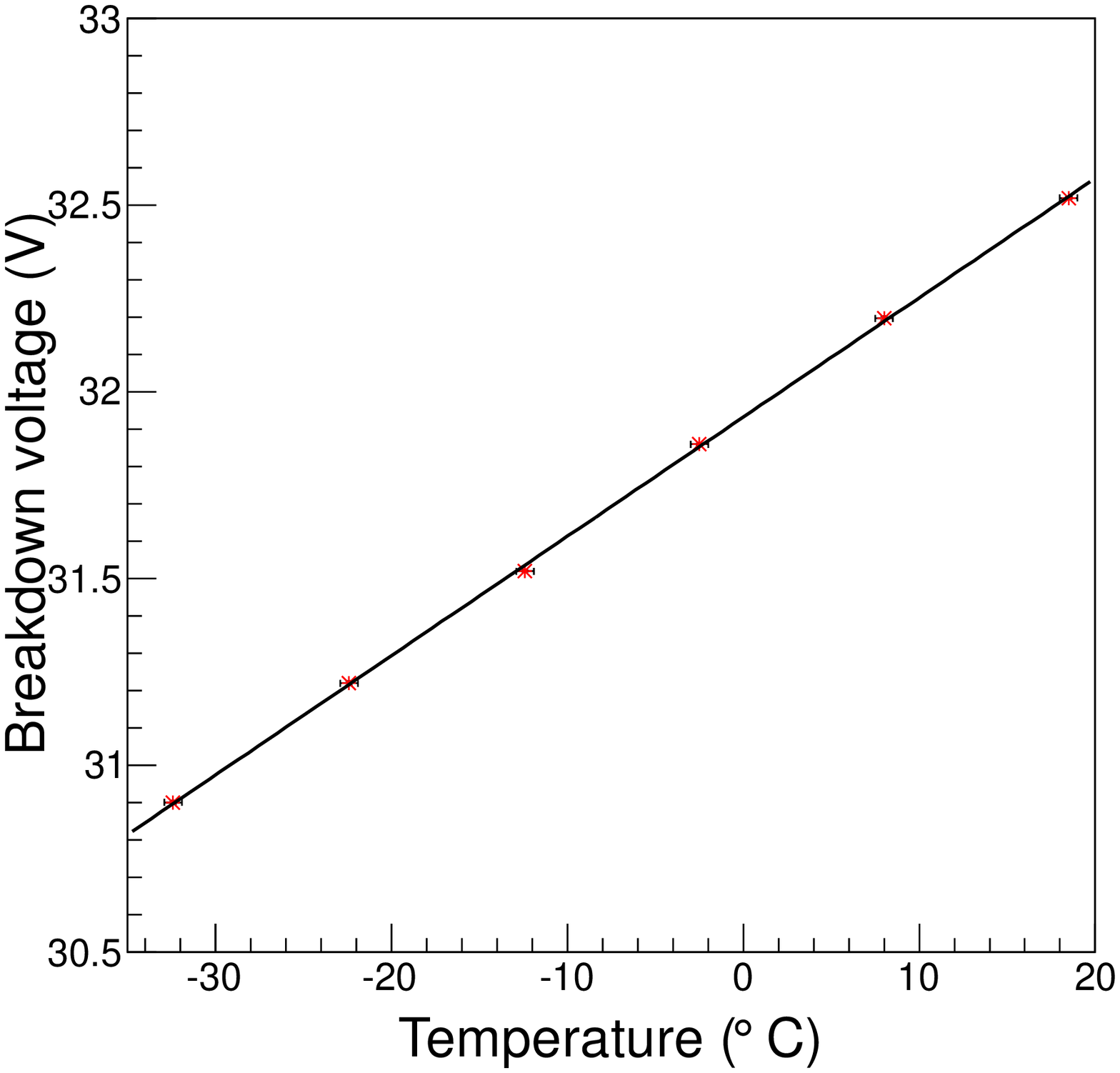}
\includegraphics[width=0.35\linewidth,clip]{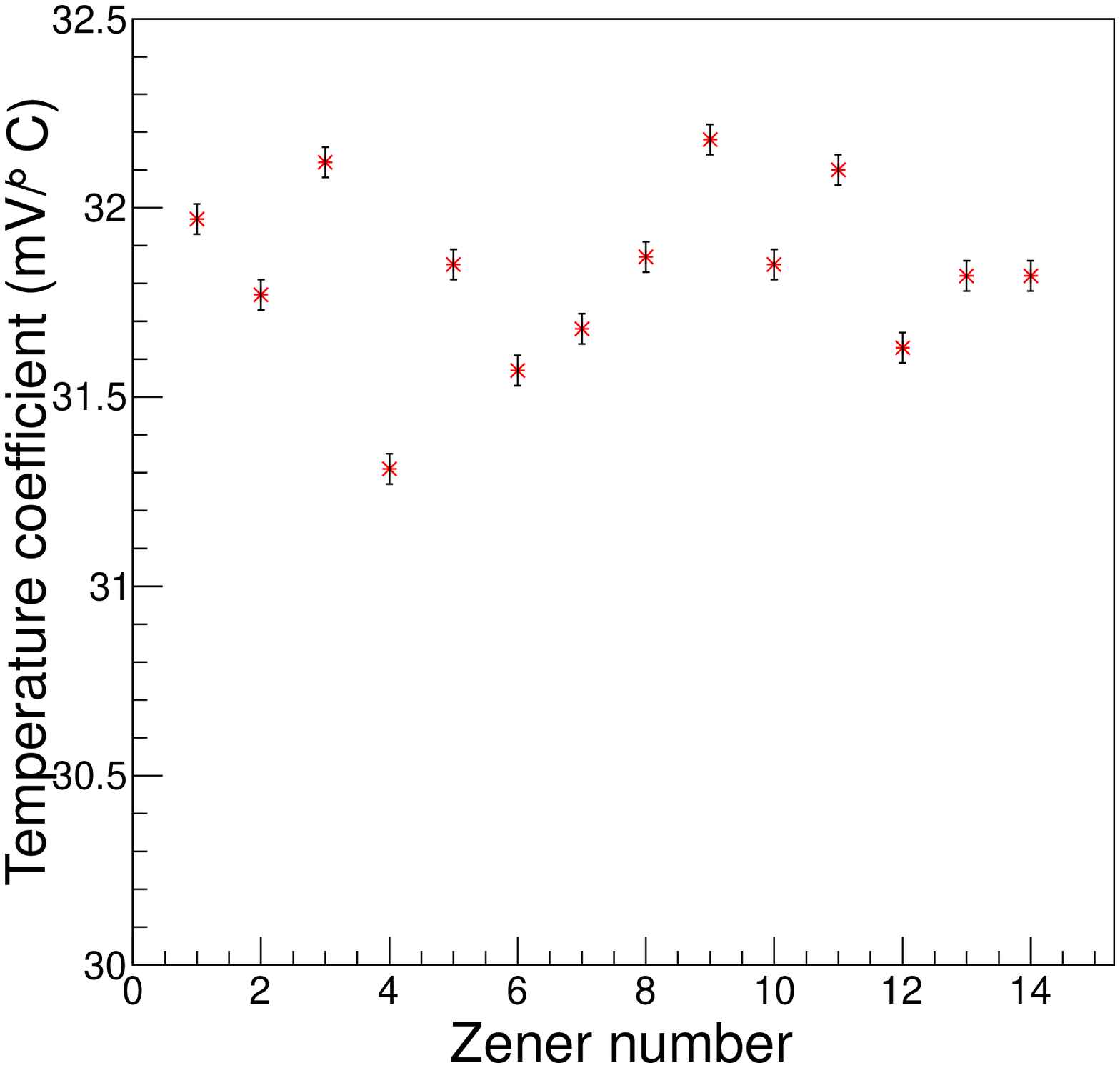}
\caption{Left panel: measured breakdown voltage dependence on the operating temperature for one BZV55-C33 zener diode sample. The solid line in the figure is the linear fitted result. Right panel: measured temperature coefficient for 14 zener diode samples.}
\label{fig.zenerbreakdown}
\end{center}
\end{figure*}
The zener diode working in the avalanche mode has a positive temperature coefficient and the temperature coefficient increases with the zener voltage. The zener diode with a zener voltage of 33~V is chosen as the temperature sensor for the analog power supply system. The temperature coefficient of the zener diode is about 32~mV$/^{\circ} \mathrm C$, so the sum of two such zener diodes is about 64~mV$/^{\circ} \mathrm C$ which nearly equals to that of the MPPC. The output voltage of the power supply system based on the two 33~V zener diodes can range from 66~V to 74~V near room temperature. The over-voltage applied to the MPPC when biased by the power supply system is in the range from 0~V to 3.5~V.  The zener diode BZV55-C33 from NXP Semiconductors was studied as the zener diode for the power supply system, as shown in Figure.~\ref{fig.voltagesource}. The breakdown voltage dependence for the 14 BZV55-C33 diode samples were studied in the temperature range from $-32.3^{\circ} \mathrm C$ to $18.5^{\circ} \mathrm C$. The measured breakdown voltage for one of the zener diodes was shown in the Figure.~\ref{fig.zenerbreakdown}. The temperature coefficient for the zener diode was derived from the breakdown voltage dependence on the operating temperature by linear fitting. The measured temperature coefficient for the zener diode samples were ranging from 31.31~mV$/^{\circ} \mathrm C$ to 32.18~mV$/^{\circ} \mathrm C$, as shown in the right panel of Figure.~\ref{fig.zenerbreakdown}. The measured breakdown voltage for the samples at $8^{\circ} \mathrm C$ are in the range from 32.06~V to 32.20~V. The difference for the breakdown voltage of different zener diode samples is within 0.14~V.

\subsection{Performance of the designed analog power supply}

\begin{figure}[htp]
\begin{center}
\includegraphics[width=1.0\linewidth,clip]{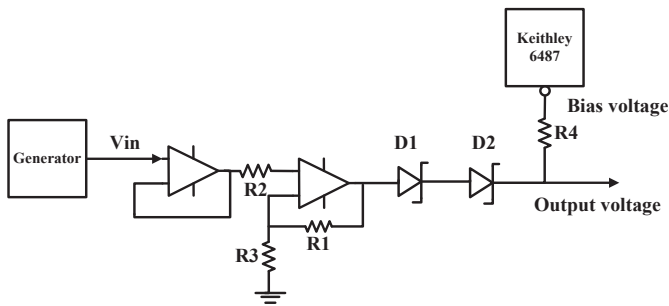}
\caption{Block diagram of experimental setup for the measurement of the temperature coefficient of the analog power supply.}
\label{fig.zenercompensation-measure}
\end{center}
\end{figure}
\begin{figure}[htp]
\begin{center}
\includegraphics[width=0.8\linewidth,clip]{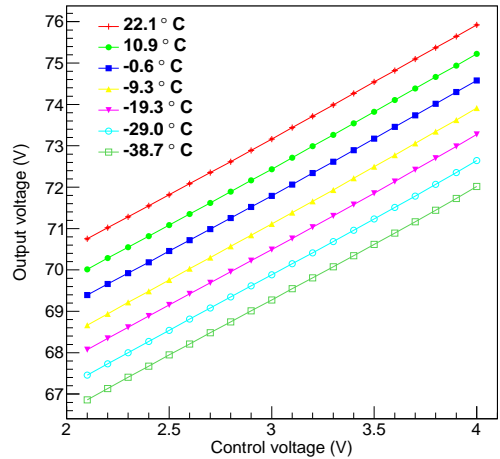}
\includegraphics[width=0.8\linewidth,clip]{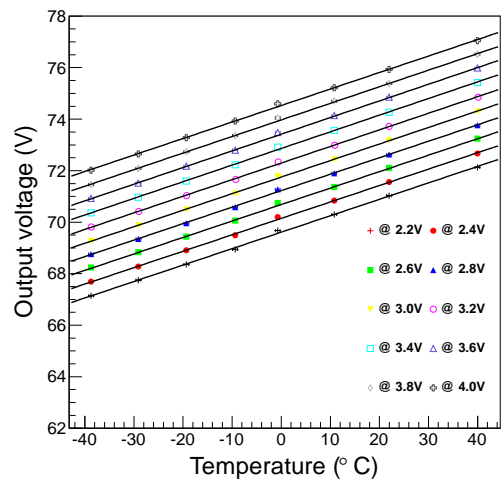}
\caption{Measured output voltage as linear function of control voltage (upper panel) and output voltage as function of operating temperature (lower panel). The solid lines in the figures show the linear fit results.}
\label{fig.outputvsinput}
\end{center}
\end{figure}

The measured temperature coefficient of the zener diode is approximately 32~mV$/^{\circ} \mathrm C$. This means that the temperature coefficient for the analog power supply system based on two zener diodes is about 64~mV$/^{\circ} \mathrm C$. This is larger than that of the MPPC. Two zener diodes with the minimum temperature coefficient were chosen as the D1 and D2 diode in the power supply system as shown in Fig.~\ref{fig.voltagesource}. The performance of the designed power supply system were measured at different environment temperatures ranging from -$38.7^{\circ} \mathrm C$ to 22.1$^{\circ} \mathrm C$. The experimental setup is shown as Figure.~\ref{fig.zenercompensation-measure}. The input voltage $V_{in}$ was provided by a generator to control the output voltage. The bias voltage applied to resistor R4 is provided by the Keithley 6487 with a voltage of 80~V. During the measurement, the Keithely 6487 measured the current ($I$) flowing through the resistor R4. The output voltage can be derived by subtracting the cross voltage applied to the resistor R4 from the 80~V bias, $V_{\mathrm output}=V_{\mathrm bias}-I \cdot R4$.

\begin{figure}[htp]
\begin{center}
\includegraphics[width=0.8\linewidth,clip]{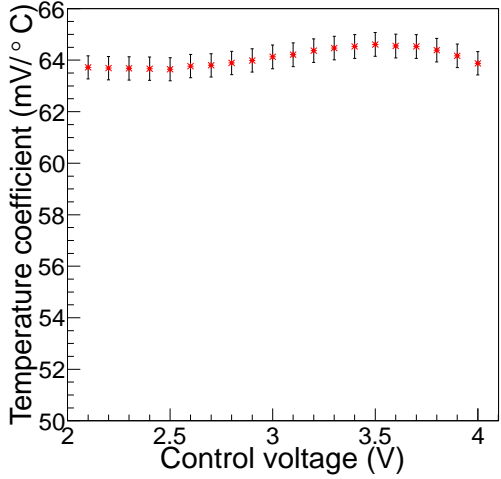}
\caption{Measured temperature coefficient at different control voltages for the power supply system.}
\label{fig.coefficient-tcs}
\end{center}
\end{figure}

The measured output voltage at different environment temperatures are shown in the upper panel of Fig.~\ref{fig.outputvsinput}. It is found that the output voltage of the power supply system increases as a linear function of the control voltage ($\mathrm V_{\mathrm in}$). The measured output voltage at room temperature of about 22.1$^{\circ} \mathrm C$ ranging from 70.75~V to 75.92~V. This means that the designed power supply system can bias the MPPC with an over-voltage ranging from 0.06~V to 5.23~V. From this measurement, the output voltage dependence on the operating temperature can be derived by fixing the control voltage at a constant value. The results are shown in the lower panel of Fig.~\ref{fig.outputvsinput}. The output voltage increases as a linear function of the operating temperature at a given control voltage. The temperature coefficient of the output voltage can be derived by fitting the output voltage dependence on the operating temperature with a linear function. The measured temperature coefficient at different control voltages were shown in the Fig.~\ref{fig.coefficient-tcs}. It is found that the measured temperature coefficient is in the range from 63.65~mV/$^{\circ} \mathrm C$ to 64.61~mV/$^{\circ} \mathrm C$. The maximum temperature coefficient was measured at the control voltage of 3.6~V. The measured temperature coefficient of the power supply system is larger than that of the MPPC by a value ranging from 4.5~mV/$^{\circ} \mathrm C$ to 5.2~mV/$^{\circ} \mathrm C$. This means that if the MPPC is powered by the designed power supply system, the gain offset caused by the variation of environment temperature will be reduced from $\sim$2.9$\%$/$^{\circ} \mathrm C$ to $\sim0.25\%$/$^{\circ} \mathrm C$ when the over-voltage is about 2~V.

\section{Gain control and stabilization of MPPC}\label{sect:performance}
The gain of the MPPC when biased by the designed power supply system for gain control was measured for comparison with that biased at 72.1~V by Keithley 6487 Picoammeter/Voltage Source without gain control. The measurement experimental setup is shown as in Fig.~\ref{fig.setup}.

\begin{figure}[htp]
\begin{center}
\includegraphics[width=0.8\linewidth,clip]{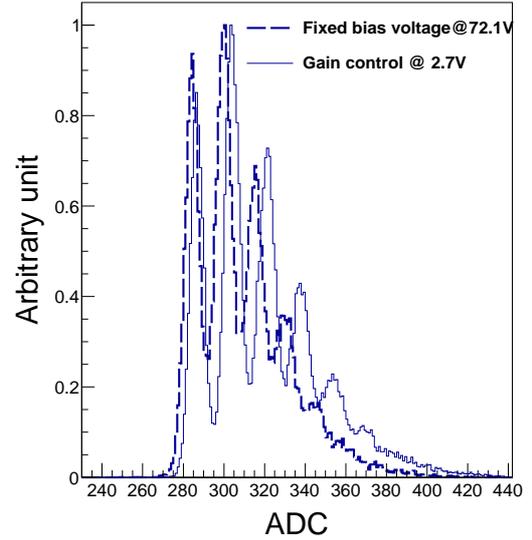}
\caption{Measured pulse-height distribution of the MPPC with and without the designed voltage source for gain control at 20.9$^{\circ} \mathrm C$.}
\label{fig.compare-noise}
\end{center}
\end{figure}

The gain was measured at different environment temperatures ranging from -42.7$^{\circ} \mathrm C$ to +20.9$^{\circ} \mathrm C$. The gain with gain control was measured with a control voltage of 2.7~V. This means that the MPPC can be biased at 72.3~V by the designed power supply system at 20.9$^{\circ} \mathrm C$. The measured pulse-height of the MPPC at 20.9$^{\circ} \mathrm C$ was shown in Fig.~\ref{fig.compare-noise}.
The equivalent noise charge (ENC) of the electronic readout system including the amplifier could be derived from the pedestal of the pulse-height distribution of MPPC. The derived ENC without and with the gain control were $1.16 \times 10^{5}$ and $1.05 \times 10^{5}$ respectively. These results showed that the designed voltage source for gain control do not induce extra noise to the whole electronic readout system.

\begin{figure}[htp]
\begin{center}
\includegraphics[width=0.8\linewidth,clip]{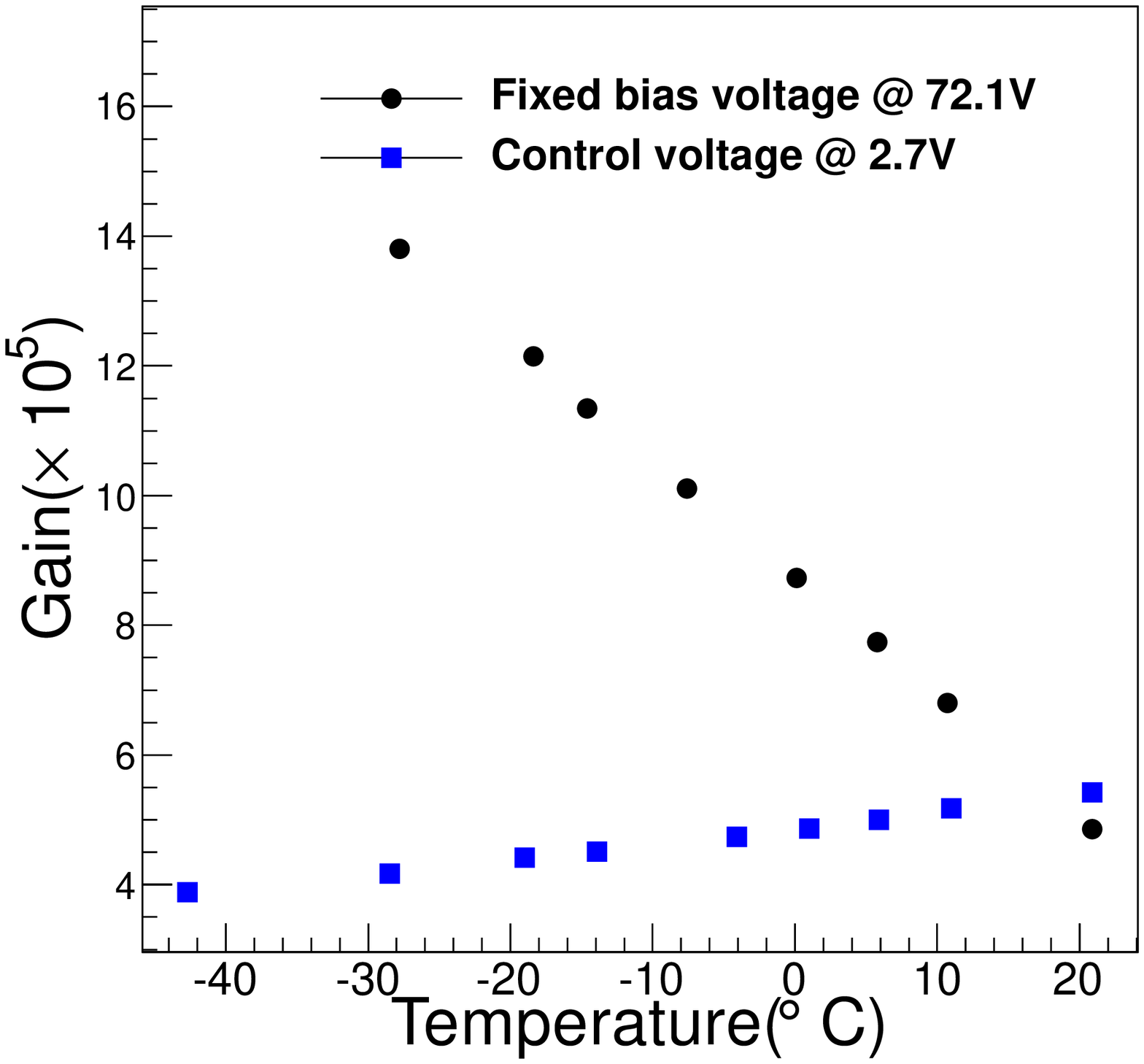}
\includegraphics[width=0.8\linewidth,clip]{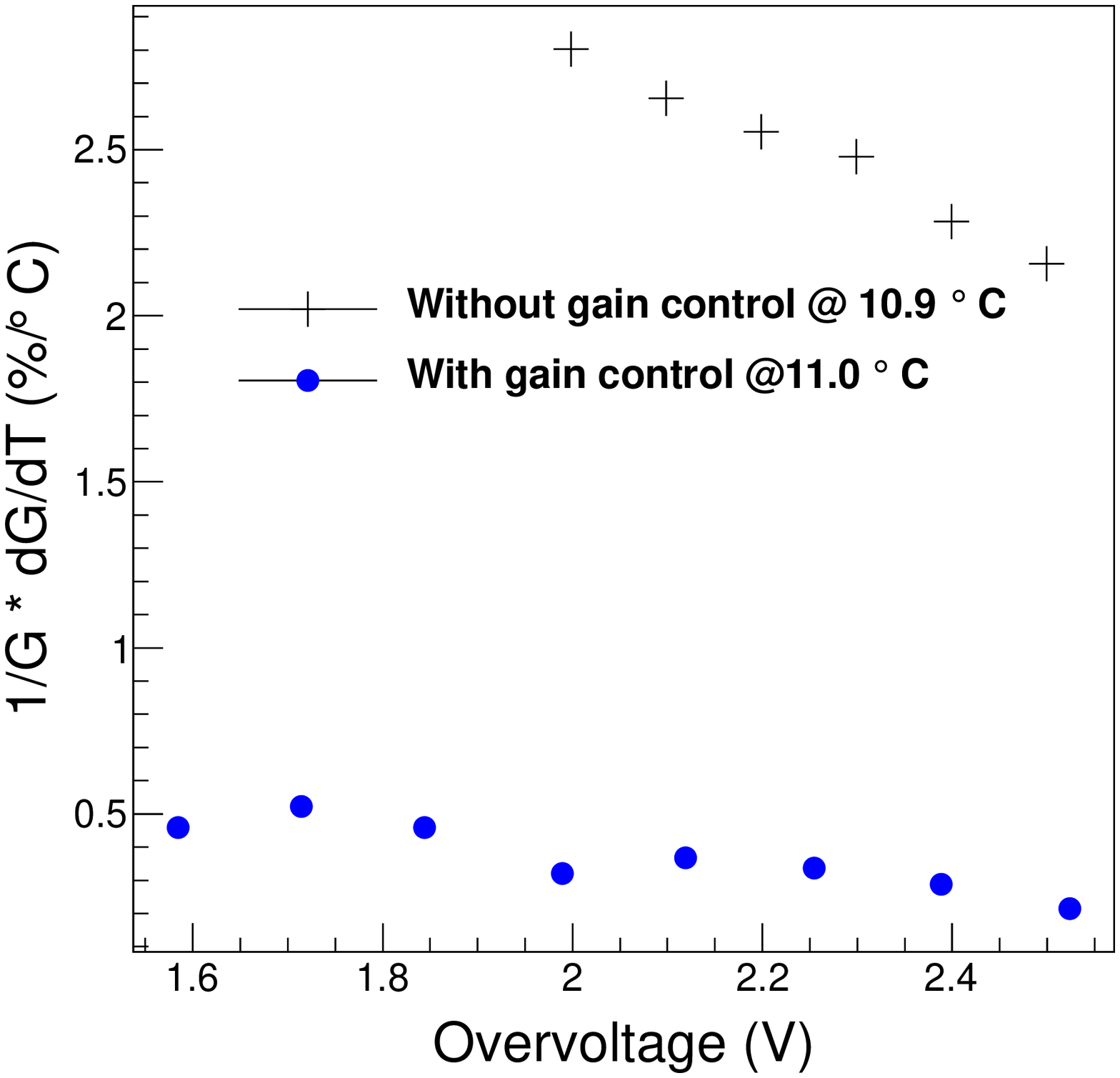}
\caption{Measured gain variance vs temperature with and without the designed voltage source for gain control.}
\label{fig.gainactivecontrol}
\end{center}
\end{figure}

The measured gain of MPPC is shown in the upper panel of Fig.~\ref{fig.gainactivecontrol}. It is found that the gain of the MPPC without gain control becomes smaller when the environment temperature increases, while the gain with gain control becomes larger. This is because the temperature coefficient of the designed power supply system is larger than that of the MPPC with a value of about 5mV/$^{\circ} \mathrm C$. The gain of the MPPC without gain control increased from $4.9 \times 10^{5}$ to $1.4 \times 10^{6}$ when temperature ranging from 20.9$^{\circ} \mathrm C$ to -42.7$^{\circ} \mathrm C$. The gain has been increased by a factor of 1.84 due to the 63.6$^{\circ} \mathrm C$ variation of the environment temperature, while the gain variation with gain control was reduced to 0.278 at the same temperature variation, ranging from $5.4 \times 10^{5}$ to $3.9 \times 10^{5}$. This means that the compensation for variation of breakdown voltage caused by the change of temperature is efficient. The gain dependence on the environment temperature can be fitted by a linear function $G(T)=A+B\cdot T$. The linear fitted results for the gain without gain control are $A = (8.741 \pm 0.017) \times 10^{5}$ and $B = -(0.183 \pm 0.001) \times 10^{5}$. The fitted results with gain control are $A = (4.883 \pm 0.013) \times 10^{5}$ and $B = (0.024 \pm 0.001) \times 10^{5}$. From the fitted results, it is found that the absolute slope of the gain dependence on operating temperature is reduced by an order of magnitude with the designed power supply for gain control.

The relative temperature coefficient for the gain of the MPPC with and without the designed power supply system for gain control was measured for comparison. The temperature coefficient for the gain is defined as following
\begin{equation}
A=|\frac{dG}{dT}|\cdot\frac{1}{G}
\end{equation}
where $G$ represents the gain of MPPC and $T$ represents the environment temperature.
The biased voltage applied to the MPPC was tuned to measure the gain at different over-voltages.   The gain was measured at different environment temperatures.
The measured temperature coefficient for the MPPC at different over-voltages is shown in the lower panel of Fig.~\ref{fig.gainactivecontrol}. It is found that the temperature coefficient for the gain is in the range from 2.8$\%$/$^{\circ} \mathrm C$ to 2.15$\%$/$^{\circ} \mathrm C$ when the MPPC was biased with an over-voltage ranging from 2~V to 2.5~V without gain control. The temperature coefficient was reduced below 0.5$\%$/$^{\circ} \mathrm C$ when the MPPC was biased by the designed power supply system for gain control. The temperature coefficient becomes smaller when the MPPC was biased with a larger over-voltage. The measured temperature coefficient for the gain of MPPC was reduced approximately by an order of magnitude when biased by the designed programmable power supply system for gain control. It was reduced from 2.8$\%$/$^{\circ} \mathrm C$ to 0.3$\%$/$^{\circ} \mathrm C$. This measured result is consistent with the prediction for the designed power supply system based on the measured temperature coefficient of the MPPC and zener diodes.

\section{Conclusions and Discussion}\label{sect:discussion}

\begin{figure}[htp]
\begin{center}
\includegraphics[width=0.7\linewidth,clip]{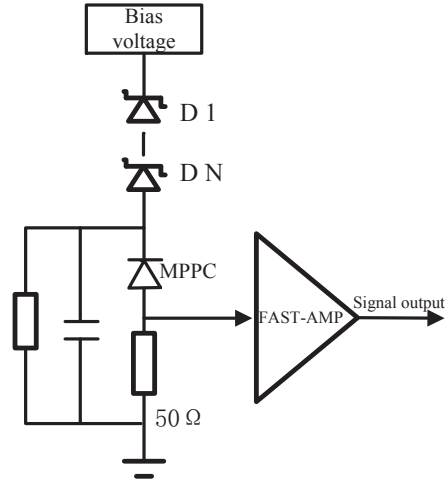}
\caption{Low power dissipation temperature compensation circuit for the MPPC.}
\label{fig.mppccompensation}
\end{center}
\end{figure}

The gain of the MPPC was measured at different operating temperatures to determine the gain temperature coefficient. The measured temperature coefficient of the breakdown voltage was $(59.4 \pm 0.4$ mV)$/^{\circ} \mathrm C$. The zener diode working in avalanche region has a similar positive linear temperature coefficient of zener voltage. By combining two zener diodes in series, an analog power supply system can be designed to have a positive linear temperature coefficient which equals that of the MPPC. The measured temperature coefficient for the designed power supply system based on two BZV55-C33 zener diodes was in the range from 63.65~mV/$^{\circ} \mathrm C$ to 64.61~mV/$^{\circ} \mathrm C$. The relative gain temperature coefficient for the MPPC was reduced from 2.8$\%$/$^{\circ} \mathrm C$ to 0.3$\%$/$^{\circ} \mathrm C$ when biased at over-voltage of 2~V by the designed power supply system. A good stabilization of the gain for varies temperature from -42.7$^{\circ} \mathrm C$ to +20.9$^{\circ} \mathrm C$ was achieved. The designed analog power supply system has been used as the bias voltage source for the MPPC on board the HXMT. Now, the flight-model of the analog power supply is under construction.

The main drawback of the designed analog power supply was the large power dissipation which was about 50~mW. This will limit the application of the designed power supply system especially in the case with thousands of MPPCs. The designed power supply system includes an operating amplifier which is quite complex for the large detector system with thousands of MPPCs. The designed power supply system can be simplified by using negative linear temperature coefficient zener diodes, just as shown in Fig.~\ref{fig.mppccompensation}. By this optimization, the power consumption is expected to be reduced from 50~mW to about 5~mW. Several zener diodes which have negative linear temperature coefficient are in series with the MPPC to compensation the voltage offset caused by the temperature variation. The zener diode is biased by a precise voltage source with feedback system to eliminate the effect of temperature. As the zener diodes have a negative temperature coefficient, the voltage that applied to the MPPC will have a positive temperature coefficient with a absolute value equals to the sum of that of the zener diodes. By using proper number of zener diodes with given temperature coefficient, the temperature coefficient of the voltage applied to the MPPC can become equal to that of the MPPC. Therefore the gain of the MPPC can be maintained at a constant value at different operating temperatures. This method can be used to maintain the gain of a large array of MPPC for its simpler circuit and lower power consumption.
\section*{Acknowledgment}\label{sect:acknowledgement}

This work is supported by the HXMT project, the Space Science Advance Research Program of Chinese Academy of Sciences and the National Natural Science Foundation of China under Grant No.11603025.

\bibliographystyle{model1a-num-names}
\bibliography{references}

\begin{thebibliography}{17}
\expandafter\ifx\csname natexlab\endcsname\relax\def\natexlab#1{#1}\fi
\providecommand{\bibinfo}[2]{#2}
\ifx\xfnm\relax \def\xfnm[#1]{\unskip,\space#1}\fi
\bibitem[{{Lu} et~al.(2014)}]{Lu201430}
\bibinfo{author}{Y.~{Lu}}, et~al., \bibinfo{journal}{\nima}
  \bibinfo{volume}{743} (\bibinfo{year}{2014}) \bibinfo{pages}{30--38}.
\bibitem[{{Buzhan} et~al.(2003)}]{Buzhan2003}
\bibinfo{author}{P.~{Buzhan}}, et~al., \bibinfo{journal}{\nima}
  \bibinfo{volume}{504} (\bibinfo{year}{2003}) \bibinfo{pages}{48}.
\bibitem[{{Kovaltchouk} et~al.(2005)}]{Kovaltchouk2005}
\bibinfo{author}{V.~D. {Kovaltchouk}}, et~al., \bibinfo{journal}{\nima}
  \bibinfo{volume}{538} (\bibinfo{year}{2005}) \bibinfo{pages}{408}.
\bibitem[{{Del Guerra} et~al.(2011)}]{DelGuerra2011}
\bibinfo{author}{A.~{Del Guerra}}, et~al., \bibinfo{journal}{\nima}
  \bibinfo{volume}{648} (\bibinfo{year}{2011}) \bibinfo{pages}{232}.
\bibitem[{{Li} et~al.(2016){Li}, {Xu}, {Liu} et~al.}]{Li201663}
\bibinfo{author}{Z.~{Li}}, \bibinfo{author}{Y.~{Xu}},
  \bibinfo{author}{C.~{Liu}}, et~al., \bibinfo{journal}{\nima}
  \bibinfo{volume}{822} (\bibinfo{year}{2016}) \bibinfo{pages}{63--70}.
\bibitem[{{Jim}(2010)}]{Freeman2010393}
\bibinfo{author}{F.~{Jim}}, \bibinfo{journal}{\nima} \bibinfo{volume}{617}
  (\bibinfo{year}{2010}) \bibinfo{pages}{393--395}.
\bibitem[{{Li} et~al.(2012){Li}, {Xu}, {Liu} et~al.}]{2012lee}
\bibinfo{author}{Z.~{Li}}, \bibinfo{author}{Y.~{Xu}},
  \bibinfo{author}{C.~{Liu}}, et~al., \bibinfo{journal}{\nima}
  \bibinfo{volume}{695} (\bibinfo{year}{2012}) \bibinfo{pages}{222--225}.
\bibitem[{{Miyamoto} et~al.(2009)}]{Miyamoto2009}
\bibinfo{author}{H.~{Miyamoto}}, et~al., in: \bibinfo{booktitle}{Proceedings of
  the 31th Cosmic Ray Conference}.
\bibitem[{{Marrocchesi} et~al.(2009)}]{Marrocchesi2009}
\bibinfo{author}{P.~S. {Marrocchesi}}, et~al., \bibinfo{journal}{\nima}
  \bibinfo{volume}{602} (\bibinfo{year}{2009}) \bibinfo{pages}{391}.
\bibitem[{Dorosz et~al.(2013)}]{Dorosz2013202}
\bibinfo{author}{P.~Dorosz}, et~al., \bibinfo{journal}{\nima}
  \bibinfo{volume}{718} (\bibinfo{year}{2013}) \bibinfo{pages}{202--204}.
\bibitem[{Baszczyk et~al.(2016)}]{Baszczyk201685}
\bibinfo{author}{M.~Baszczyk}, et~al., \bibinfo{journal}{\nima}
  \bibinfo{volume}{824} (\bibinfo{year}{2016}) \bibinfo{pages}{85--86}.
\bibitem[{Shukla et~al.(2016)}]{Shukla}
\bibinfo{author}{R.~A. Shukla}, et~al., \bibinfo{journal}{Review of Scientific
  Instruments} \bibinfo{volume}{87} (\bibinfo{year}{2016}).
\bibitem[{{Yamamoto} et~al.(2011)}]{Yamamoto2011}
\bibinfo{author}{S.~{Yamamoto}}, et~al., \bibinfo{journal}{Physics in Medicine
  and Biology} \bibinfo{volume}{56} (\bibinfo{year}{2011})
  \bibinfo{pages}{2873--2882}.
\bibitem[{{Licciulli} et~al.(2013)}]{Licciulli2013}
\bibinfo{author}{F.~{Licciulli}}, et~al., \bibinfo{journal}{Ieee Transactions
  on Nuclear Science} \bibinfo{volume}{60} (\bibinfo{year}{2013})
  \bibinfo{pages}{606--611}.
\bibitem[{{Bencardino} and {Eberhardt}(2009)}]{Bencardino2009}
\bibinfo{author}{R.~{Bencardino}}, \bibinfo{author}{J.~E. {Eberhardt}},
  \bibinfo{journal}{Ieee Transactions on Nuclear Science} \bibinfo{volume}{56}
  (\bibinfo{year}{2009}) \bibinfo{pages}{1129--1134}.
\bibitem[{{Musienko}(2009)}]{Musienko2009a}
\bibinfo{author}{Y.~{Musienko}}, \bibinfo{journal}{\nima} \bibinfo{volume}{598}
  (\bibinfo{year}{2009}) \bibinfo{pages}{213}.
\bibitem[{{Danilov}(2009)}]{Danilov2009}
\bibinfo{author}{M.~{Danilov}}, \bibinfo{journal}{\nima} \bibinfo{volume}{604}
  (\bibinfo{year}{2009}) \bibinfo{pages}{183}.

\end{thebibliography}






\end{document}